\documentclass{emulateapj}
\usepackage{color}

\slugcomment{Accepted 2008 March 19}

\shorttitle{Lack of O-line excess in the Coma outskirts}
\shortauthors{Takei et al.}

\newcommand{\NH}{N_\mathrm{H}}

\newcommand{\LU}{\mathrm{LU}}

\begin{document}
\title{
On the lack of strong O-line excess in the Coma cluster
outskirts from {\it Suzaku}}

\email{Y.Takei@sron.nl}

\author{
Yoh Takei\altaffilmark{1}, 
Eric D. Miller \altaffilmark{2}, 
Joel N. Bregman\altaffilmark{3},
Shunsuke Kimura\altaffilmark{4}, 
Takaya Ohashi\altaffilmark{5},
Kazuhisa Mitsuda\altaffilmark{4}, 
Takayuki Tamura\altaffilmark{4}, 
Noriko Y. Yamasaki\altaffilmark{4}, 
Ryuichi Fujimoto\altaffilmark{6}
}

\altaffiltext{1}{SRON Netherlands Institute for Space Research,
Sorbonnelaan 2, 3584CA, Utrecht, The Netherlands}
\altaffiltext{2}{Massachusetts Institute of Technology,
Kavli Institute for Astrophysics and Space Science,
77 Massachusetts Ave 37-551, Cambridge, MA 02139}
\altaffiltext{3}{
University of Michigan, Department of Astronomy, 
500 Church St., Ann Arbor, MI 48109
}
\altaffiltext{4}{
Institute of Space and Astronautical Science (ISAS),
Japan Aerospace Exploration Agency (JAXA), 
3-1-1 Yoshinodai, Sagamihara, Kanagawa, 229-8510, Japan}
\altaffiltext{5}{
Department of Physics, Tokyo Metropolitan University,
1-1 Minami-Osawa, Hachioji, Tokyo 192-0397, Japan
}
\altaffiltext{6}{
Department of Physics, Kanazawa University, Kakuma-machi,
Kanazawa 920-1192, Japan
}

\begin{abstract}
About half of the baryons in the local Universe are thought to reside in
the so-called warm-hot intergalactic medium (WHIM) at temperatures of
0.1-10 million K.  Thermal soft excess emission in the spectrum of some
cluster outskirts that contains OVII and/or OVIII emission lines is
regarded as evidence of the WHIM, although the
origin of the lines
is controversial due to strong Galactic and solar system
foreground emission.
We observed the Coma-11 field, where
the most prominent thermal soft excess has ever been reported,
with Suzaku XIS in order to make clear the origin of the excess.  
We did not confirm OVII or OVIII excess emission.  
The OVII and OVIII intensity in
Coma-11 is more than 5$\sigma$ below that reported before and we
obtained $2\sigma$ upper limits of
2.8 and 2.9~$\mathrm{photons~ cm^{-2}~ s^{-1}~ sr^{-1}}$
for \ion{O}{7} and \ion{O}{8},
respectively.
The intensities are consistent with those 
in another field (Coma-7) that we measured,
and with other measurements in the Coma outskirts
(Coma-7 and X Com fields with XMM-Newton).
We did not confirm the
spatial variation within Coma outskirts.
The strong
oxygen emission lines previously
reported are likely due to solar wind charge
exchange.
\end{abstract}


\section{Introduction}

Based on N-body simulations of cosmological
large-scale structure formation
\citep[e.g.,][]{2006ApJ...650..560C}, 
a significant (30--50\%) fraction of baryons are thought to reside 
in a `warm-hot' gaseous phase ($T=10^{5-7}$~K), which is hard to
detect with the instruments currently in operation.  This warm-hot
gas, whose density is $10^{-6}$--$10^{-4}\mathrm{~cm^{-3}}$, is called
the warm-hot intergalactic medium (WHIM).  Firm detection of the
WHIM is important because it is the most promising candidate for the
``missing baryons''; i.e.,  it is thought to explain the discrepancy
between the baryon density observed in the local universe
\citep{fukugita98:_cosmic_baryon_budget} and that in the
distant universe
\citep{rauch97:_opacit_ly_fores_implic_omega} or that calculated from
the observed fluctuations of the cosmic microwave background
\citep{2007ApJS..170..377S}.
Although $\sim10\%$ of the baryons has been resolved as the WHIM
in the temperature range $T=10^{5-6}$~K through
\ion{O}{6} absorption features observed with
{\it FUSE} and {\it HST} \citep{2007arXiv0709.4030D},
more baryons are
thought to reside in a hotter phase at $T=10^{6-7}$~K.  
The WHIM of
this temperature may be detected via emission or absorption lines from
highly ionized elements in X-ray spectra, such as \ion{O}{7} and
\ion{O}{8}.

Despite many observations so far, the
existence of the WHIM in the hotter phase is not yet confirmed
\citep[see review of][and references therein]{2007ARA&A..45..221B}.
Studies of the WHIM using absorption lines only lead to
one disputed detection toward Mkn~421
\citep{2005Natur.433..495N}, a result
which was not confirmed later and was thus questioned
\citep{2007ApJ...656..129R,2006ApJ...652..189K}.
Possible detections of emission from the WHIM 
have been reported in cluster outskirts, as a soft X-ray 
($E\lesssim 0.5$~keV) excess containing \ion{O}{7} and \ion{O}{8}
lines
\citep{2003A&A...397..445K,finoguenov03:_xmm_newton_x_coma}.
The origin of O line emission is, however, controversial
 due to strong Galactic and
solar system foreground emission, which has a spectral shape similar
to the WHIM
\citep{2005A&A...443...29B,2006ApJ...644..167B}.
In particular, X-rays induced by solar wind charge exchange (SWCX)
could produce serious contamination because of its high variability
\citep[e.g.,][]{2000ApJ...532L.153C,2007PASJ...59S.133F}.
Note that there has been no detection of redshifted emission lines
from the WHIM so far.

The most prominent thermal soft excess ever reported is in the Coma
outskirts, especially in the Coma-11 field 
\citep[][hereafter FBH03]{finoguenov03:_xmm_newton_x_coma}. 
FBH03 also observed spatial variation within the Coma
outskirts; Coma-7 shows roughly five
 times smaller intensity than Coma-11.
Not only O lines, but also an EUV excess \citep{2003ApJ...585..722B}
and marginal detection of \ion{Ne}{9} emission and absorption lines
\citep[][hereafter T07]{2007ApJ...655..831T}
are reported in the Coma outskirts. These results
may support the existence of the WHIM.

We observed the Coma-11 and Coma-7 fields with the XIS
\citep{2007PASJ...59S..23K}
onboard {\it Suzaku} \citep{2007PASJ...59S...1M},
in order to make clear the origin of the excess.
The XIS is an ideal instrument to study the emission lines from
the WHIM, because it 
does not show a large
low-energy tail in the pulse-height distribution function, even in the
very soft ($E\lesssim 0.5~\mathrm{keV}$) band.  

We assume a Hubble constant of 70 km s$^{-1}$ Mpc$^{-1}$ or $h_{70}=1$
and $\Omega_\mathrm{m} = 0.3$, $\Omega_\mathrm{\Lambda} = 0.7$.
The solar metal abundance is given by \citet{2003ApJ...591.1220L}.
Unless otherwise
stated, errors and upper limits
are at the 68~\% confidence level in the figures, and at
the 90~\% confidence level in the text and tables.
The emission line intensities are quoted in line units (LU) defined
as $\mathrm{photons~ cm^{-2}~ s^{-1}~ sr^{-1}}$.
One LU corresponds to
$8.46\times10^{-8}~\mathrm{photons~ cm^{-2}~ s^{-1}~ arcmin^{-2}}$.

\section{Observations and data reduction}

\begin{deluxetable}{lccc}
\tablecaption{Observation Information\label{tab-obsinfo}}
\tablewidth{0pt}
\tablehead{
& \colhead{Coordinates }
& \colhead{Exposure\tablenotemark{a}}
\\
& \colhead{(J2000.0)} & &
} 
\startdata
   Coma-11 & 
 (12$^{\rm h}$58$^{\rm m}$31.7$^{\rm s}$, 28$^\circ$24$'$13.2$''$) 
& 53.02 ks 
\\
   Coma-7 & 
(12$^{\rm h}$57$^{\rm m}$22.9$^{\rm s}$, 28$^\circ$08$'$58.1$''$) 
      & 25.26 ks 
\\
  ComaBKG & 
(13$^{\rm h}$15$^{\rm m}$00.0$^{\rm s}$, 31$^\circ$39$'$28.0$''$) 
	      & 30.78 ks 
\enddata
\tablenotetext{a}{After screening as described in the text}
\end{deluxetable}

We observed three fields from 2007 Jun 19 to 2007 Jun 22;
two of them are in the Coma outskirts, 
namely Coma-11 and Coma-7
that FBH03 
observed with {\it XMM-Newton}, and 
the other is an offset observation (hereafter we call
it ComaBKG) of 5.0-degree distance
from the Coma center (NGC 4874).
The coordinates and net exposure times 
are summarized in Table~\ref{tab-obsinfo}.
The XIS was operated in the normal mode and with the
spaced-row charge injection (SCI).
We analysed the clean-event data of 
Ver~2.0 processing with standard parameters,
but with the latest gain calibration (Ver~2.6.1.16).
We excised the last 6.7 ks of the Coma-7 observation, during which
the XIS suffered a problem with zero-level determination.

We excluded the regions where point sources are found or 
which are irradiated by the $^{55}$Fe calibration sources.
The point sources were identified
using the 
XMM-Newton Serendipitous Source Catalogue
\citep[2XMM;][]{watsonCatalogue}, as well as
by examining the XIS images by eyes.
Response files were created using Suzaku ftools xisrmfgen 
version 2007-05-14 and xissimarfgen version 2007-09-22
with CALDB released on 2008-01-14.
We assumed uniform emission from the sky.
The spectra and response files of the two XIS-FIs (XIS0 and 3) are
added using ftools mathpha, marfrmf, and addrmf.
X-ray spectra are analysed with XSPEC~12.4.0.

\section{Spectral analysis}

\subsection{ComaBKG field}

ComaBKG is observed in order to precisely determine the foreground
and background emission, in particular 
O and Ne line intensities around the Coma cluster.
The instrumental background,
estimated using a ftool xisnxbgen version 2007-11-23 
with COR2 as the sorting parameter \citep{2008PASJ...60S..11T},
were subtracted from the ComaBKG spectra.

\begin{figure}
\epsscale{0.95}
\plotone{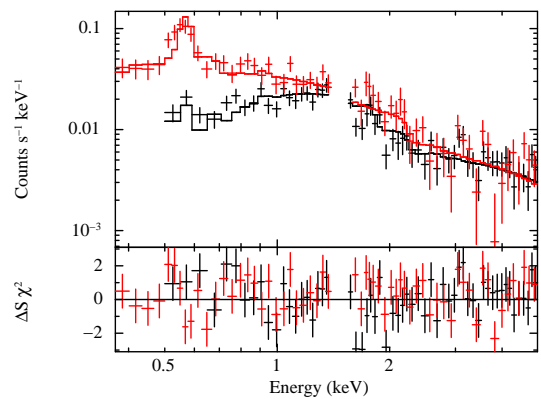}
\caption{Spectra and best-fit models (top)
and residuals in units of sigma (bottom) of ComaBKG.  
Red and black show BI and FI, respectively.
Instrumental background is subtracted.
\label{fig-bkgspectra}}
\end{figure}

We modeled the spectra with a sum of two thin thermal
plasma models (APEC in XSPEC) and a power-law to represent emission from
the local hot bubble (LHB), Milky-Way halo (MWH) and extragalactic
cosmic X-ray background (CXB), respectively.  MWH and CXB were
convolved with absorption in the Galaxy (WABS in XSPEC).
We fitted the spectra 
in the energy range 0.5--5.0~keV for the
XIS-FI and 0.35--5.0~keV for XIS1 (XIS-BI). 
The spectra of XIS-FI and XIS-BI were fitted simultaneously.
We ignored 1.40--1.55~keV
where a relatively large uncertainty exists in the instrument background
\citep{2008PASJ...60S..11T}
The redshift and abundance of the two thermal plasma models were fixed
to 0 and 1.0 solar, respectively.  The
temperature of the LHB was also fixed to 0.07~keV, while the temperature
of the MWH was allowed to vary.
The column density of the Galactic hydrogen gas was fixed to
$1.0\times10^{20}~\mathrm{cm^{-2}}$ \citep{2005A&A...440..775K}.
Figure~\ref{fig-bkgspectra} and Table~\ref{tab-bestfitBKG} show
the best-fit model.

\begin{deluxetable}{llc}
\tablecaption{Best-fit parameters of ComaBKG.\label{tab-bestfitBKG}}
\tablewidth{0pt}
\tablehead{
\colhead{Parameter} & \colhead{Unit}
& \colhead{Value} 
} 
\startdata
Galactic $N_H$ & $\mathrm{cm^{-2}}$ & 
 $1.0\times10^{20}$ (fixed) \\
LHB temperature & keV & 0.07 (fixed) \\
LHB emission measure& $\mathrm{pc~cm^{-6}}$& 
 $(3.0\pm3.0)\times10^{-2}$\\
MWH temperature & keV & $0.144^{+0.021}_{-0.027}$\\
MWH emission measure & $\mathrm{pc~cm^{-6}}$& 
 $(8.5^{+7.8}_{-3.0})\times10^{-3}$\\
CXB photon index & & $1.52\pm0.09$\\
 CXB normalization\tablenotemark{a} &
 & $7.7\pm0.5$\\
$\chi^2/dof$ thermal model& & 153.40/113=1.36\\
\tableline
\ion{O}{7} intensity & LU &
 $6.5\pm1.0$\\
\ion{O}{8} intensity & LU &
 $1.1^{+0.9}_{-0.5}$ \\
\ion{Ne}{9} intensity & LU &
 $0.15^{+0.15}_{-0.14}$\\
\ion{Ne}{10} intensity & LU &
 $0~(<0.2)$ \\
$\chi^2/dof$ with lines& & 140.22/105=1.34
\enddata
\tablenotetext{a}{
In a unit of
$\mathrm{photons~cm^{-2}~s^{-1}~keV^{-1}~sr^{-1}}$
 at 1~keV}
\end{deluxetable}

We also fitted the spectra with another model in order to
determine \ion{O}{7}, \ion{O}{8}, \ion{Ne}{9} and \ion{Ne}{10}
intensity more precisely.
Four Gaussians were added
to represent 
the four line emissions, while O and Ne
abundance of the LHB and MWH were fixed to zero.
The energies of the Gaussians were allowed to vary
$\pm35~\mathrm{eV}$ around the resonance line energy
(0.574~keV for \ion{O}{7}, 0.653~keV for \ion{O}{8},
0.922~keV for \ion{Ne}{9} and 1.022~keV for \ion{Ne}{10})
in order to compensate the uncertainty in the energy-scale calibration,
while we fixed the line width to zero.
The intensities of the four lines 
are also shown in Table~\ref{tab-bestfitBKG}.
The relatively large $\chi^2$ value may be due to incorrect
calibration of the energy resolution.  The $\chi^2$
improved to 128.18 
when we added 50~eV (FWHM) additional width
into the model with lines.
The best-fit parameters were not sensitive 
to this calibration uncertainty;
they stay the same within the error at 90\% confidence level.
We confirmed that the fitting is not improved when we allow $\NH$ to vary.

\begin{deluxetable*}{llcc}
\tablecaption{Best-fit parameters of Coma11 and Coma7.
\label{tab-fitspectraComa11-7}}
\tablewidth{0pt}
\tablehead{
\colhead{Parameter} & \colhead{Unit}
& \colhead{Coma-11} & \colhead{Coma-7}
} 
\startdata
Galactic $\NH$ & $\mathrm{cm^{-2}}$ & 
 $1.0\times10^{20}$ (fixed) &
 $1.0\times10^{20}$ (fixed) \\
 Temperature & keV & $6.7^{+0.4}_{-0.3}$ & $8.3^{+0.8}_{-0.6}$\\
Abundance & solar & $0.36\pm0.11$ & $0.21\pm0.15$\\
Redshift &  & 0.0231 (fixed) & 0.0231 (fixed) \\
EM & $\mathrm{pc~cm^{-6}}$& 
 $0.100\pm0.002$ & $0.110\pm0.002$\\
$\chi^2/dof$ without lines & & 307.51/275=1.12 & 295.40/281=1.05\\
\tableline
\ion{O}{7} intensity & LU &
 $0.0~(<0.9)$ & $0.2~(<2.1)$\\
\ion{O}{8} intensity & LU &
 $1.1\pm0.8$ & $1.7\pm1.1$\\
\ion{Ne}{9} intensity & LU &
 $0.1~(<0.4)$ & $0.4~(<0.8)$\\
\ion{Ne}{10} intensity & LU &
 $0.3\pm0.2$ & $0.3~(<0.6)$ \\
$\chi^2/dof$ with lines & & 297.71/271=1.10 & 284.92/277=1.03
\enddata
\end{deluxetable*}

\begin{figure*}[htb]
\epsscale{.95}
\begin{center}
\plottwo{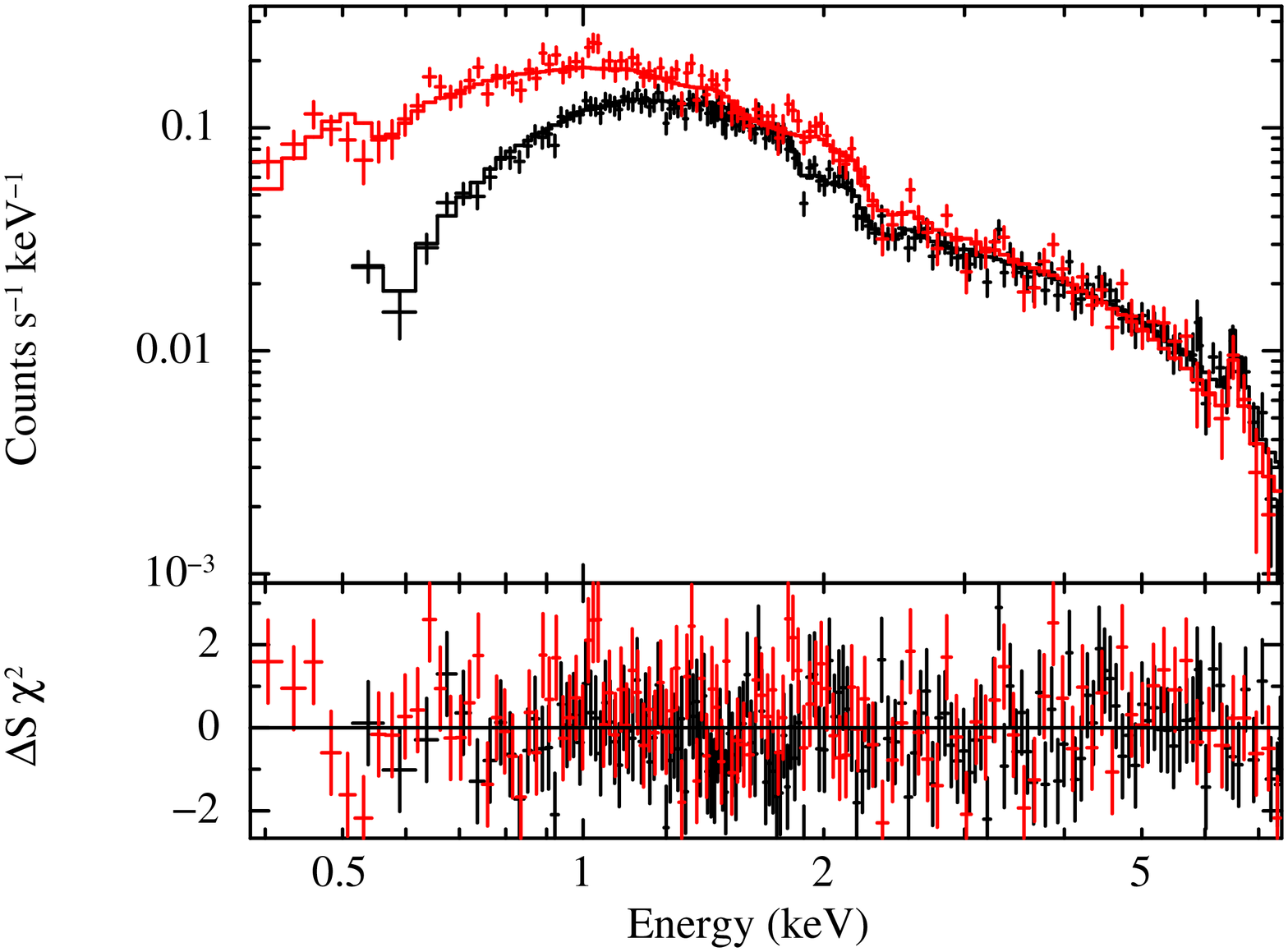}{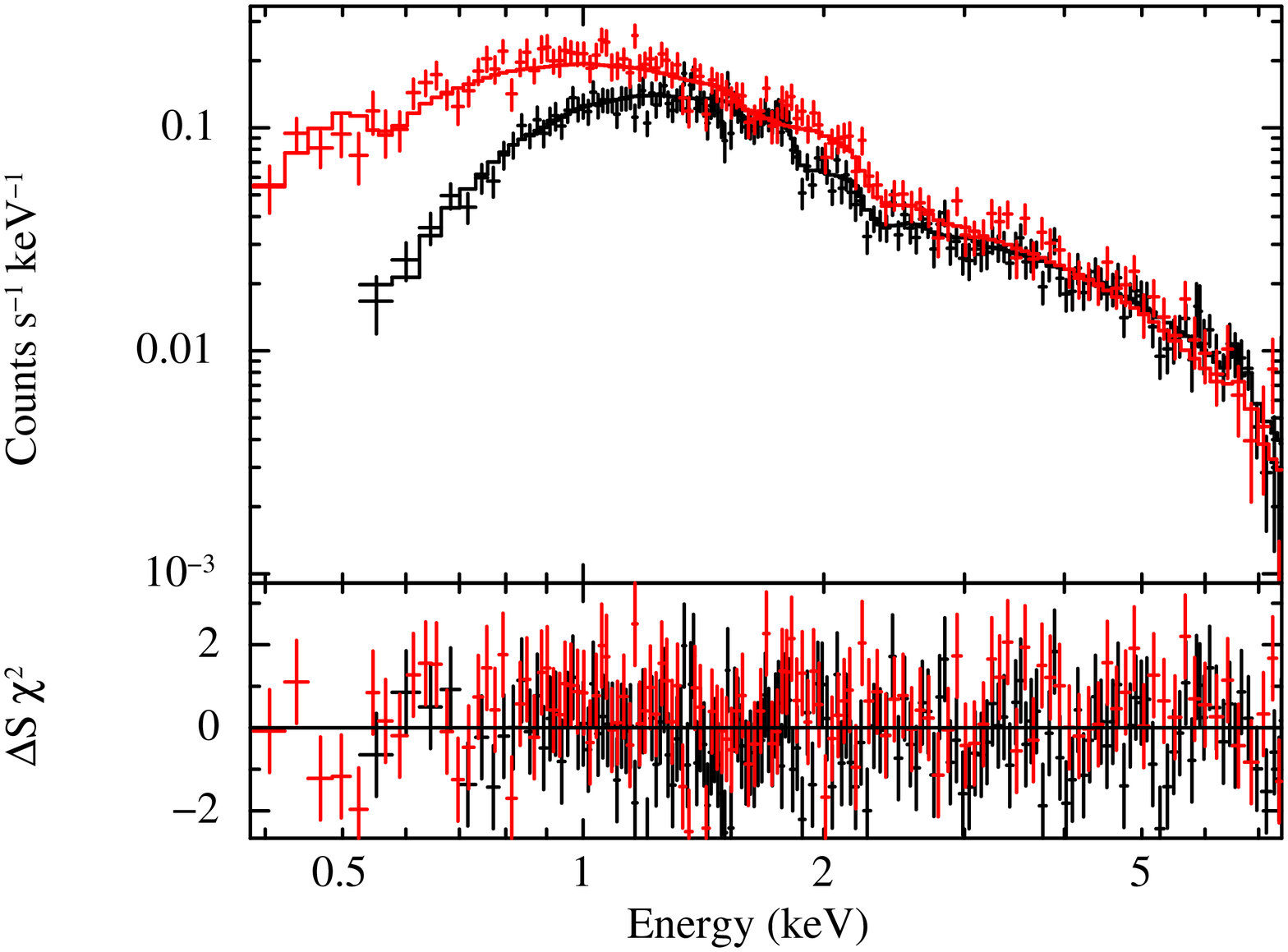}
\caption{Spectra of Coma-11 (left) and Coma-7 (right).
Red and black data correspond to BI and FI respectively.
ComaBKG spectra were subtracted to correct the background emission.
}
\label{fig-spectra-woexcess}
\end{center}
\end{figure*}

\subsection{Coma-11 and Coma-7 fields}

The back- and foreground emission of Coma-11 and Coma-7
were corrected by subtracting
the ComaBKG spectra that were extracted
from the same region as Coma-11
or Coma-7 in the detector coordinates.
We then fitted the spectra 
0.5--7.5~keV for XIS-FI and 0.35--7.5~keV for XIS-BI)
by the emission 
of the intracluster medium (ICM), i.e.,
a thin thermal plasma model (APEC) convolved with Galactic absorption (wabs).
The spectra of XIS-FI and XIS-BI were again fitted together.
The spectra and the best-fit models are shown in
Figure~\ref{fig-spectra-woexcess}, and the best-fit parameters are in
Table~\ref{tab-fitspectraComa11-7}.
We did not see a soft excess, i.e., emission from the WHIM.
Next, we added four Gaussians to the model,
corresponding to lines of \ion{O}{7},
\ion{O}{8}, \ion{Ne}{9} and \ion{Ne}{10} at the cluster redshift,
in order to investigate possible contributions of the WHIM.
The central energy of each line was fixed to that of the
resonance lines at redshift $z=0.0231$, and the width of the line was
fixed to zero.
The results are shown in Table~\ref{tab-fitspectraComa11-7}.
The fit is slightly improved ($\Delta \chi^2\sim10$) by
adding the lines.  However, the significance of each line is
less than $3\sigma$.  
The fitting without FI also gave 
the same results within the statistical errors, which excludes
the possible systematic bias due to the low statistics of the
FI data below 1~keV.
We also tried to fit the spectra with
two thin thermal plasma models, i.e., one for the intracluster
medium and one for the WHIM.  This did not improve the fit.
To obtain robust upper limits of line intensities,
we consider systematic uncertainties in calibration:
the centroid energy ($\pm35$~eV),
the energy resolution (10--80~eV FWHM),  
energy scale ($\pm50$~eV),
and the contamination thickness 
($\pm5\times10^{-17}~ \mathrm{cm^{-2}}$).
The $2\sigma$ upper limits we obtained
for \ion{O}{7}, \ion{O}{8},
\ion{Ne}{9}, and \ion{Ne}{10} are
1.4, 2.7, 0.5, and 0.9~LU for Coma-11 and 
2.6, 3.7, 1.2, and 1.1~LU for Coma-7, respectively.
The uncertainty in Galactic O intensity
should also be taken into account. 
From the variation among background fields in T07, we adopted
$\pm1.2$~LU for \ion{O}{7} and $\pm0.6$~LU for \ion{O}{8} as $1\sigma$.
Then, \ion{O}{7} and \ion{O}{8} $2\sigma$ upper limits are
2.8 and 2.9~LU for Coma-11 and 
3.5 and 3.9~LU for Coma-7, respectively.

\section{Discussion}

\begin{figure*}[htb]
\epsscale{.95}
\begin{center}
\plottwo{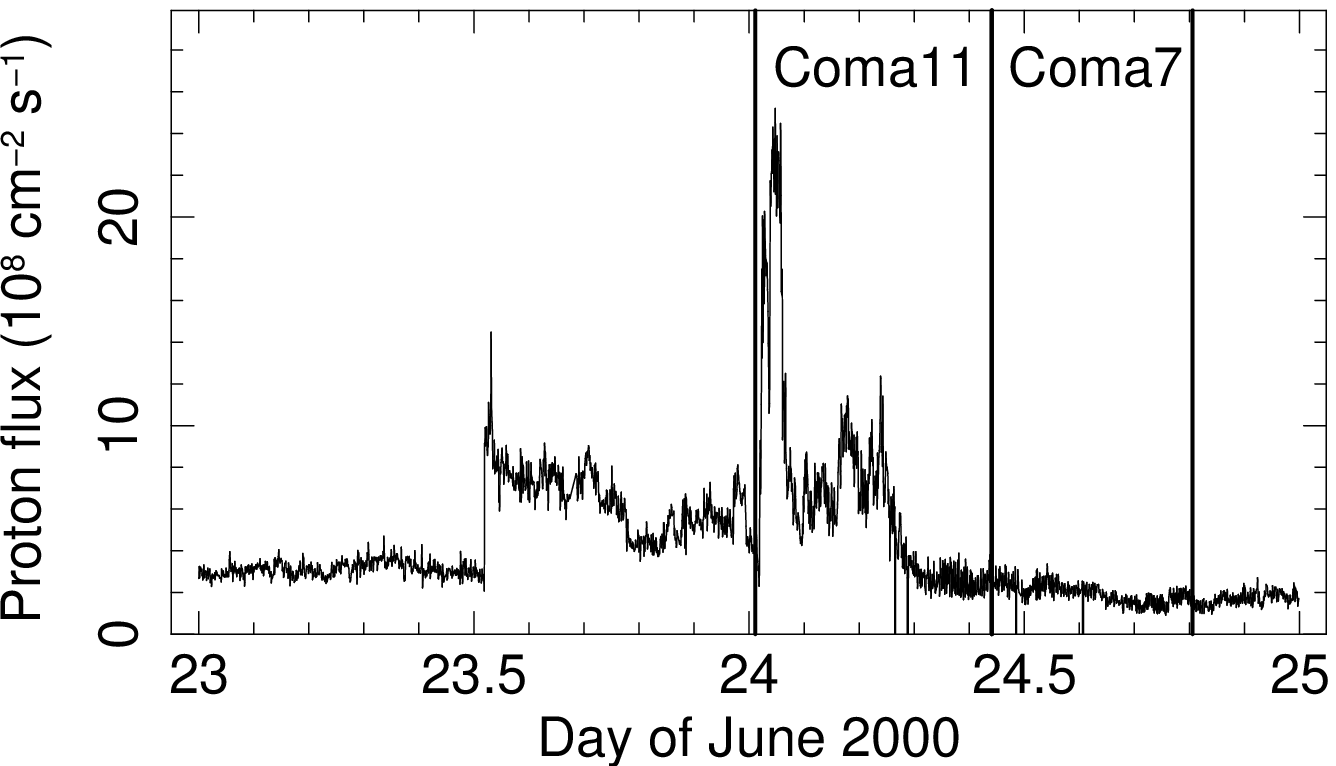}{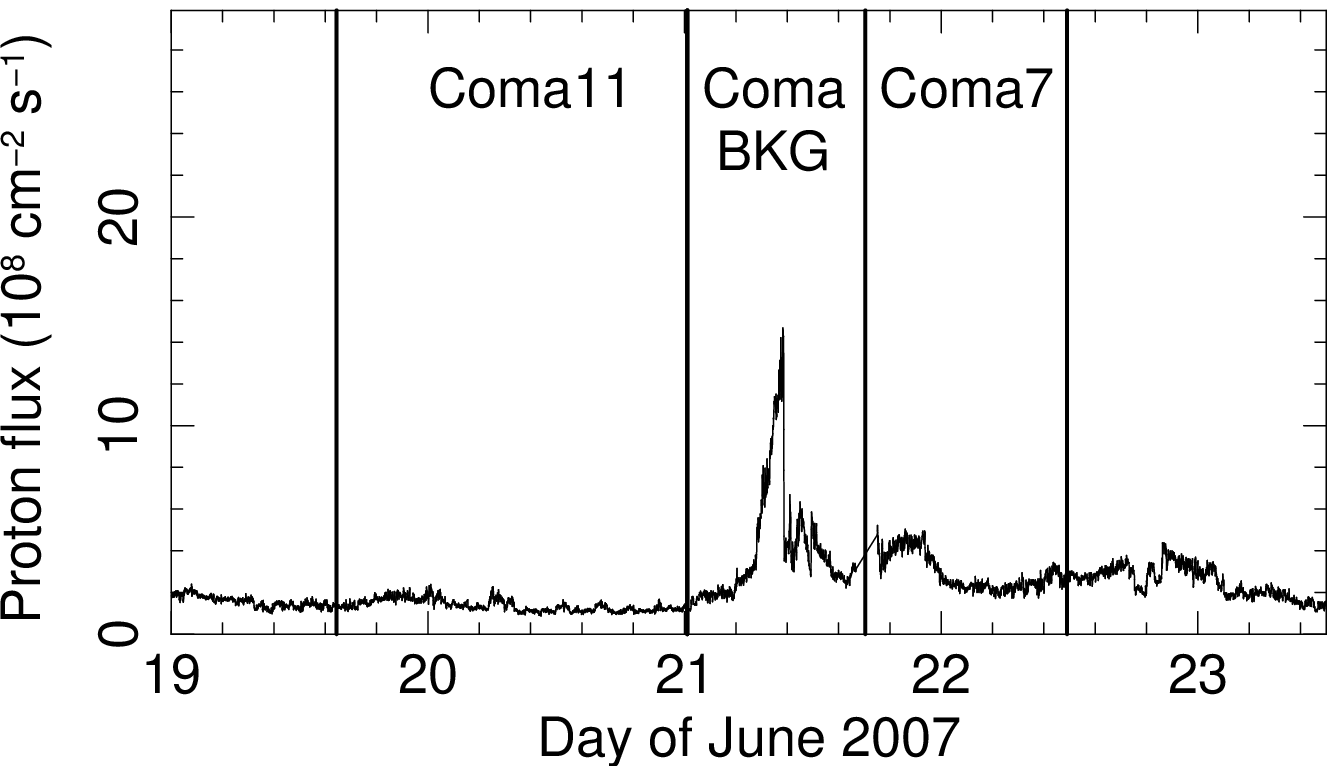}
\caption{Solar wind proton flux as a function of time,
around Coma-11 and Coma-7 observations with {\it XMM-Newton} (left)
and around our observations with {\it Suzaku} (right),
measured with ACE SWEPAM instrument.
The time of {\it XMM-Newton} and {\it Suzaku} observations
is indicated with vertical lines.
}
\label{fig-solarwind}
\end{center}
\end{figure*}

We measured the spectra of Coma-11 and Coma-7 with
the Suzaku XIS.
The spectra were 
 fitted with a single thin thermal plasma model and the emission
from the WHIM was not confirmed.
Here we compare our results with previous 
{\it XMM-Newton} observations.
We do not discuss the lower energy excess ($E<0.5$~keV)
observed with ROSAT and EUVE,
because the XIS is not sensitive in the energy.

FBH03 observed Coma-11 and Coma-7 with {\it XMM-Newton}.
The \ion{O}{7} and \ion{O}{8} intensity measured in FBH03 are
$6.4\pm0.8$ and $6.2\pm0.8$ LU in Coma-11, and
$1.4\pm0.8$ and $0.9\pm0.6$ LU in Coma-7, respectively.
The intensities of Coma-7 in FBH03
are consistent with our {\it Suzaku} observation,
while those of Coma-11 are discrepant at the $5\sigma$ level.
The best-fit ICM temperature and abundance
in Coma-11 are also inconsistent.
Even if we fixed the temperature and abundance to the
best-fit values of FBH03,
the discrepancies of the O line intensities remain as large.
The O intensities observed with {\it Suzaku} are
similar to those of Coma-7; we did not confirm the spatial
variation.

T07 measured \ion{O}{7}, \ion{O}{8}
and \ion{Ne}{9} emission line intensities with 
{\it XMM-Newton} in the field
of X Com and four background fields whose distance from
NGC~4874 is 46.2--162.7~arcmin.
The intensities of \ion{O}{7}, \ion{O}{8} and \ion{Ne}{9}
lines measured by T07 are
$6.6\pm0.2$,
$1.8\pm0.2$,
and $0.7\pm0.1$~LU in the X Com field and 
$7.4\pm1.2$,
$2.6\pm0.6$,
and $0.4\pm0.1~\LU$,
in the (averaged) background fields, respectively.
The values of the X Com field shown above
contain contributions of the background emission.
The intensities of the X Com field are consistent within 
90\% confidence range with
those of the Coma-11 and Coma-7 measured in this work,
and those of the background field of T07 are also consistent
with those of ComaBKG.

The origin of the large discrepancy in Coma-11
O intensities between this work and FBH03 could be
contamination by SWCX emission.
The solar wind proton flux 
observed with 
ACE (Advanced Composition Explorer\footnote{
http://www.srl.caltech.edu/ACE/ASC}) 
showed flares
both during Coma-11 observation 
of FBH03 (2.5$\times10^{9}~\mathrm{cm^{-2}~s^{-1}}$) and
during ComaBKG observation of this work
(1.5$\times10^{9}~\mathrm{cm^{-2}~s^{-1}}$)
as shown in Figure~\ref{fig-solarwind}.
Both are $\sim8$ times larger than the quiescent state
around each observation time.
Hence, two possibilities should be considered:
O intensities of Coma-11 in FBH03 are
overestimated, or those of ComaBKG in this work are
overestimated and thus those of Coma-11 in this work are
underestimated.
However, the latter case is not likely because the O-line
intensities
are consistent with or even smaller than T07 measurements of
background fields around the Coma cluster.
There is no indication of the geocoronal SWCX emission; 
the first point in the line of sight (LOS)
where the geomagnetic field started open
was always higher than 10 times Earth 
radius\footnote{
Calculated using the software GEOPACK-2005 and T96 magnetic field model
({\tt
http://geo.phys.spbu.ru/$\sim$tsyganenko/modeling.html
}),
with CDAWeb solar-wind parameters
({\tt http://cdaweb.gsfc.nasa.gov/cdaweb/sp\_phys/})
}
\citep{2007PASJ...59S.133F,2005JGRA..11003208T},
and there was no variation in the \ion{O}{7} and 
\ion{O}{8} flux during the ComaBKG observation.
The heliospheric SWCX emission could be weak since
it is faint at solar minimum
on the high ecliptic lattitude line of sight
\citep{2007A&A...475..901K}.
Moreover, even in the case that
the ComaBKG observation was heavily
contaminated with the SWCX emission, 
the $2\sigma$ upper limits of Coma-11
\ion{O}{7} and \ion{O}{8}
intensity would be no larger than
4.8~LU and 3.5~LU, respectively, which
are still lower than values of FBH03.
These values are calculated by assuming
the lower limits of \ion{O}{7} and \ion{O}{8}
intensities as the LHB contribution
without any MWH contribution,
measured toward a molecular cloud MBM~12
\citep[3.5~LU and 0~LU, respectively;][]{2007PASJ...59S.141S}.
On the other hand,
it is more likely that the FBH03 observation suffered
heliospheric SWCX contamination, since
a similar solar wind flare during {\it
XMM-Newton} observation of the Hubble deep field north
increased the \ion{O}{7} and \ion{O}{8} flux by $6.25\pm0.67$
and $5.53\pm2.87$ LU, respectively 
\citep{2004ApJ...610.1182S,2007A&A...475..901K}.
The solar cycle at the observations,
the position of the Earth and the LOS,
which also
affects the X-ray flux induced by SWCX, are similar 
between the two observations.
Note that the observation of FBH03 was performed at solar
maximum, while our observation was at solar minimum.
The field-to-field variation of \ion{O}{7} and \ion{O}{8}
flux reported by FBH03
might also be due to the variation of the SWCX flux.

\acknowledgments
YT and SK thank T. Hagihara for his instruction on
the calculation of the geomagnetic field distribution.
YT is financially supported by the Dutch Organization for Scientific
Research (NWO).
EDM and JNB gratefully acknowledge support for Suzaku activities from
NASA.

\end{document}